\documentclass[journal]{IEEEtran}
\usepackage{cite}
\usepackage{amsmath,amssymb,amsfonts}

\usepackage{algorithmic}
\usepackage{graphicx}
\usepackage{textcomp}
\usepackage{xcolor}
\usepackage{array,multirow}
\usepackage{rotating}
\usepackage{ifpdf}
\usepackage{cleveref}
\usepackage{float}
\usepackage[utf8]{inputenc}
\usepackage[T1]{fontenc}
\usepackage{array,multirow}
\usepackage{tabularx,booktabs,lipsum,graphicx}
\usepackage{rotating}
\usepackage{enumitem}
\usepackage{scalerel}
\usepackage{stfloats}
\usepackage{tikz}
\usetikzlibrary{svg.path}
\begin{document}
	\bstctlcite{IEEEexample:BSTcontrol}
	\bibliographystyle{IEEEtran}
	\title{Voltage Instability Prediction Using a Deep Recurrent Neural Network}
	%
	%
	%
	
	\author{Hannes~Hagmar,~\IEEEmembership{Student Member,~IEEE,}
		Lang Tong,~\IEEEmembership{Fellow,~IEEE,}
		Robert~Eriksson,~\IEEEmembership{Senior Member,~IEEE,}
		and Le~Anh~Tuan,~\IEEEmembership{Member,~IEEE}
		
		\thanks{
			The work presented in this paper has been financially supported by Energimyndigheten (Swedish Energy Agency) and Svenska kraftnät (Swedish National Grid) within the SamspEL program with project number 44358-1. The work of Lang Tong was supported in part by the U.S. National Science Foundation under grant 1932501 and 1809830.}
	}
	
	%
	%

	\markboth{}%
	{Shell \MakeLowercase{\textit{et al.}}: Bare Demo of IEEEtran.cls for IEEE Journals}
	%
	
	
	\maketitle
	
	\begin{abstract}
		
		This paper develops a new method for voltage instability prediction using a recurrent neural network with long short-term memory. The method is aimed to be used as a supplementary warning system for system operators, capable of assessing whether the current state will cause voltage instability issues several minutes into the future. The proposed method use a long sequence-based network, where both real-time and historic data are used to enhance the classification accuracy. The network is trained and tested on the Nordic32 test system, where combinations of different operating conditions and contingency scenarios are generated using time-domain simulations. The method shows that almost all N-1 contingency test cases were predicted correctly, and N-1-1 contingency test cases were predicted with over 93 \% accuracy only seconds after a disturbance. Further, the impact of sequence length is examined, showing that the proposed long sequenced-based method provides significantly better classification accuracy than both a feedforward neural network and a network using a shorter sequence.

	\end{abstract}
	

	\IEEEpeerreviewmaketitle

	\section{Introduction}
	\IEEEPARstart{V}{oltage} instability is one of the main limitations for secure operation of a modern power system \cite{IEEE2004}. A voltage instability event can often be deceiving, where the system may seem stable for several minutes after a disturbance, only ending up in an unstable state within a short time period \cite{Glavic2011}. When instability finally is detected, the system may already have become severely degraded and the risks of an extended blackout may have increased significantly.
	
	To ensure a secure operation, system operators often use an approach called dynamic security assessment (DSA). DSA includes time-domain analysis to test the power system's dynamic response after a set of contingencies \cite{Konstantelos2017}. Assessment of the dynamic stability is a complex task, and even with recent progress in high performance computing, it is generally not feasible to assess the dynamic stability in real-time \cite{Konstantelos2017}. 
	
	To overcome this issue, various machine learning (ML) methods have been proposed in the literature. The main advantage of using ML is that high-cost computations can be performed off-line. Once the ML algorithm is trained, it can almost instantaneously provide estimations and warnings to operators that otherwise would require time-consuming computations. Examples of DSA methods based on ML are found in \cite{Cutsem1993,Mansour1997,Sun2007,Khoshkhoo2014,Konstantelos2017,Liu2018}, where mainly various decision tree (DT) or neural network (NN) methods are utilized.
	
	Voltage security assessment (VSA) is a branch in DSA that specifically examines the impact of voltage instability events. This paper deals with the \textit{emergency} applications of VSA, where the \textit{current} system state is assessed. Here, the stability of the system is not tested with respect to a set of contingencies; rather, the system may already have suffered a disturbance. The aim of these methods is to perform voltage instability prediction (VIP), allowing system operators to trigger fast remedial actions. The emergency applications of VSA using ML have been less examined in the literature, but examples include DT \cite{Cutsem1993,Diao2009,Khoshkhoo2011} and NN \cite{Hagmar2019} methods. 
	
	Previously developed methods for VIP have all in common that only instantaneous measurements are used as inputs to the VIP algorithms. These inputs represent the "state signal" that the ML algorithm use to predict the future state. Ideally, the state signal should summarize all relevant information required to determine the future state of the system. A state signal achieving this is said to have Markov property \cite{Sutton2015}. However, the dynamic response of a power system cannot be modeled as a first order Markov process using only the static states provided by available measurements in the power system. Rather, the future state of the system also depends on a range of unknown state variables such as the rotor speed of generators, tap positions, or rotor slips of induction motors. 
	
	In response to these limitations, we propose a new method based on a recurrent neural network (RNN) with long short-term memory (LSTM). LSTM networks excel at capturing long-term dependencies \cite{Schmidhuber2017}, which is an inherent aspect in long-term voltage stability \cite{Glavic2011}. The method is, to the authors' knowledge, the first of its kind to use current \textit{and} past data with the aim to enhance the available state signal and implicitly take into account unknown state variables. 
	
	The main contributions of this paper are the following: 
	
	\begin{itemize}
		\item A methodology for VIP using an LSTM network is developed. The LSTM network can utilize previous measurements, such as the trend of bus voltage magnitudes, tap changes, or fault locations, to improve the accuracy for VIP. The performance using the sequence-based approach is compared with an LSTM network using shorter sequence and a conventional NN. 
		\item A new training approach is developed to provide operators with an \textit{online} assessment tool for potential voltage instability. As time progresses after a voltage instability event, the network is capable of incorporating new observations and continuously updating the assessment.
		\item A methodology for including consecutive contingencies ($N$-$1$-$1$) into the training data is presented. The paper also examines the ability of the LSTM network to generalize for VIP under $N$-$1$-$1$ contingencies. Such ability is especially valuable in overcoming the combinatorial increase of complexity in training.
	\end{itemize}
	
	The rest of the paper is organized as follows. In Section~\ref{sec:RNN}, the theory regarding RNNs and LSTM is presented. In Section \ref{sec:Method}, the proposed method is presented along with the steps for developing the training data and the training of the LSTM network. In Section \ref{sec:Simulations}, the results and discussion are presented. Concluding remarks are presented in Section \ref{sec:Conclusions}.
	
	\section{Long short-term memory networks}
	\label{sec:RNN}
	Neural networks is a class of machine learning algorithms, highly capable of accurately approximating nonlinear functions, mapping a set of inputs to a corresponding set of target values. RNNs represent a specific type of NNs adapted for processing \textit{sequential} input data \cite{Rumelhart1986}. However, the standard implementation of RNN has difficulties in capturing long-term dependencies of events that are significantly separated in time. In an LSTM network, such information can be propagated through time within an internal state memory cell, making the network capable of memorizing features of significance~\cite{Hochreiter1997}. 
	
\begin{figure}
		\begin{center}
			\vspace{-0.6cm}
			\includegraphics[width=6.6cm]{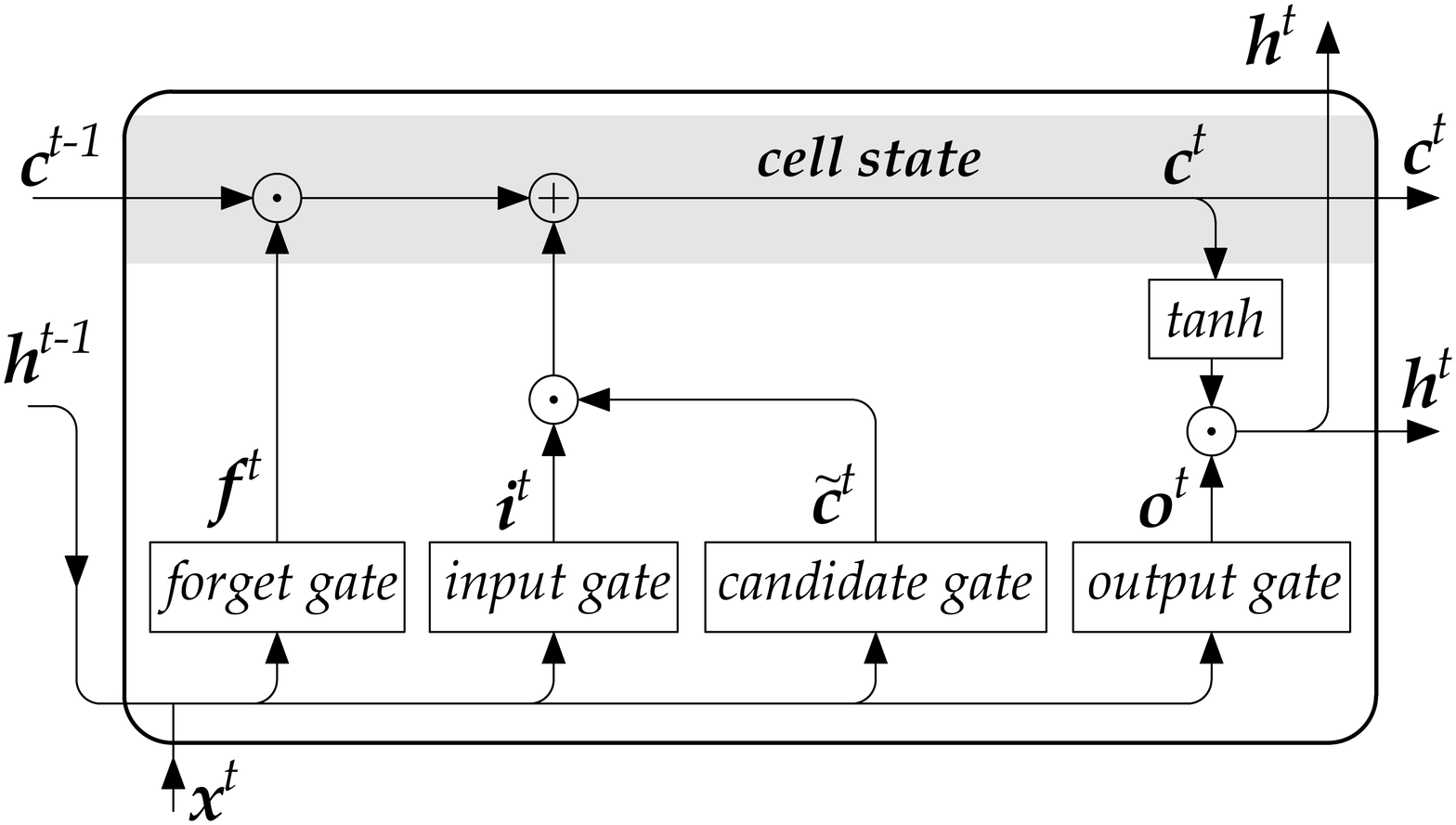}
			\vspace{-0.8cm}
			\caption{Detailed schematics of an LSTM block} 
			\label{fig:LSTMExplainedSimple}
			\vspace{-0.7cm}
		\end{center}
	\end{figure}

	A typical LSTM-block is illustrated in Fig \ref{fig:LSTMExplainedSimple}. The state memory cell, illustrated by the light grey area, is controlled by nonlinear gating units that regulate the flow in and out of the cell \cite{Schmidhuber2017}. Following \cite{Hochreiter1997} and \cite{Schmidhuber2017}, the forward operation of an LSTM block is summarized below. It should be noted that each block consists of a number of hidden LSTM cells. Vector notation is used, meaning that, for instance, the hidden state vector $\boldsymbol{h}^t$ is not the output of a single LSTM-cell at time $t$, but the output of a vector of $N$ LSTM-cells. 
	The operation of an LSTM block at a time $t$ may then be summarized by:

	\begin{equation} \label{Fgate}
	\begin{array}{l}
	\boldsymbol{f}^t = \sigma\left(\boldsymbol{W}_f\boldsymbol{x}^{t}+\boldsymbol{U}_f\boldsymbol{h}^{t-1}+\boldsymbol{b}_f\right)
	\end{array}
	\end{equation}
	\begin{equation} \label{Igate}
	\begin{array}{l}
	\boldsymbol{i}^t = \sigma\left(\boldsymbol{W}_i\boldsymbol{x}^{t}+\boldsymbol{U}_i\boldsymbol{h}^{t-1}+\boldsymbol{b}_i\right)
	\end{array}
	\end{equation}
	\begin{equation} \label{Cgate}
	\begin{array}{l}
	\tilde{\boldsymbol{c}}^t = \tanh\left(\boldsymbol{W}_c\boldsymbol{x}^{t}+\boldsymbol{U}_c\boldsymbol{h}^{t-1}+\boldsymbol{b}_c\right)
	\end{array}
	\end{equation}
	\begin{equation} \label{multi}
	\begin{array}{l}
	\boldsymbol{c}^t = \boldsymbol{f}^t\odot\boldsymbol{c}^{t-1}+\boldsymbol{i}^t\odot\tilde{\boldsymbol{c}}^t
	\end{array}
	\end{equation}
	\begin{equation} \label{Ogate}
	\begin{array}{l}
	\boldsymbol{o}^t = \sigma\left(\boldsymbol{W}_o\boldsymbol{x}^{t}+\boldsymbol{U}_o\boldsymbol{h}^{t-1}+\boldsymbol{b}_o\right)
	\end{array}
	\end{equation} 
	\begin{equation} \label{output}
	\begin{array}{l}
	\boldsymbol{h}^t = \boldsymbol{o}^t\odot\tanh(\boldsymbol{c}^t),
	\end{array}
	\end{equation}  
	where element-wise multiplication is denoted by $\odot$, $\sigma$ is the logistic sigmoid function, $\tanh$ is the hyperbolic tangent function, and with the following variables:
	
	\begin{itemize}
		\item $\boldsymbol{x}^t\in\mathbb{R}^M$: input vector to an LSTM block
		\item $\boldsymbol{h}^t$,$\boldsymbol{h}^{t-1}
		\in\mathbb{R}^N$: output vector at time $t$ respectively $t$-$1$
		\item $\boldsymbol{f}^t\in\mathbb{R}^N$: activation vector of the forget gate
		\item $\boldsymbol{i}^t\in\mathbb{R}^N$: activation vector of the input gate
		\item $\tilde{\boldsymbol{c}}^t\in\mathbb{R}^N$: vector of the the candidate gate
		\item $\boldsymbol{c}^t\in\mathbb{R}^N$: cell state memory vector
		\item $\boldsymbol{i}^t\in\mathbb{R}^N$: activation vector of the output gate
	\end{itemize}
	where $\boldsymbol{W}$, $\boldsymbol{U}$, and $\boldsymbol{b}$ represents the weight matrices and bias vectors for each gate. The superscripts $M$ and $N$ refer to the number of inputs and hidden LSTM cells in each LSTM block, respectively. 
	
	By the operation of \eqref{Fgate}, the forget gate controls what information should be stored from the previous memory cell state, and what can be discarded as irrelevant. The input gate and candidate gate control and update the memory cell state with new information by \eqref{Igate}~--~\eqref{multi}. Equations \eqref{Ogate}~--~\eqref{output} shows how the hidden state is updated by the operation of the output gate, modulated by the updated cell state memory vector.
	
\begin{figure}[H]
		\begin{center}
			\vspace{-2.1cm}
			\includegraphics[width=7.0cm]{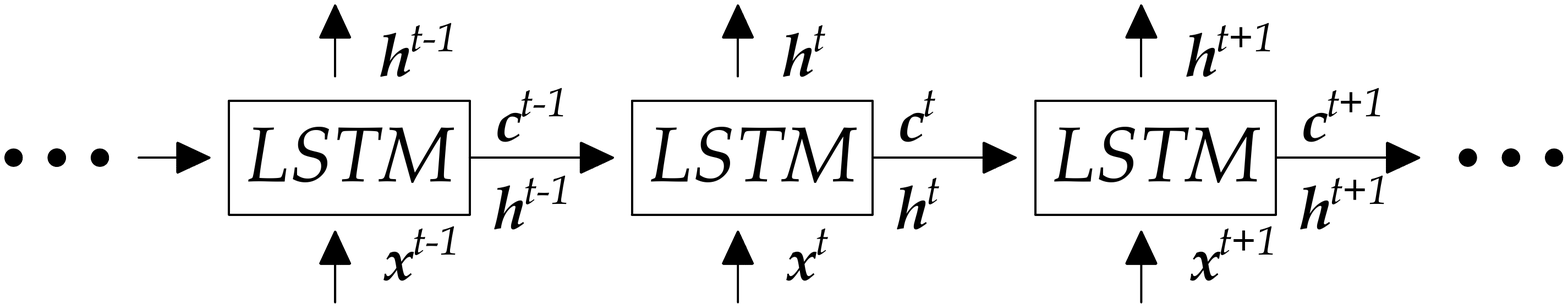}
			\vspace{-2.1cm}
			\caption{An LSTM sequence with a directed connection between the blocks} 
			\label{fig:RNN_simple}
			\vspace{-0.5cm}
		\end{center}
	\end{figure}

	An LSTM network may then be constructed by creating a sequences of several LSTM blocks. A partition of an LSTM sequence is illustrated in Fig. \ref{fig:RNN_simple}, where each block has a directed connection to the following block in the sequence. If the block is the first one in the sequence, the past system state is initialized with a preset value. For a deep LSTM network, with several stacked layers, the inputs to the deeper layers consists of the hidden states of LSTM blocks of previous layers. The cell state memory is only passed along the time sequence between LSTM blocks of the same layer. Typically, for classification purposes, an output vector $\boldsymbol{y}$ is generated by applying a nonlinear function of the hidden state implemented by a separate feedforward NN. Depending on the application of the network, output vectors may be computed for a single, or for several, LSTM block's hidden states.

	The LSTM network can then be trained using a supervised approach, where a set of training sequences and an optimization algorithm are used to update and learn suitable values for the weights matrices and bias vector parameters.
	
	\section{Methodology}
	\label{sec:Method}
	
	The proposed method for real-time VIP is based on off-line training of an LSTM network on a large data set consisting of time-domain simulation responses following a set of credible contingencies. The method is aimed to be used as a supplementary warning system that can assess the current state of the system in real time. The LSTM network takes real-time and historic measurements and attempts to assess whether the \textit{current} state will cause voltage stability issues several minutes into the future. As time progresses and if new events occur in the system, the network updates the assessment continuously. The network is also adapted to be able to indicate \textit{where} in the system instability emerges, following the approach developed in \cite{Hagmar2019}, allowing more cost-effective countermeasures.
	
	The first step of the method is the off-line generation of credible operating conditions (OCs) and contingency scenarios using time-domain simulations. The method is tested on the Nordic32 test system with all data and models as presented in \cite{IEEETESTSYSTEM}. After a representative training set is generated, training of the LSTM network is performed. Each step in the methodology is described in the following subsections. 
	
	\begin{figure}
		\begin{center}\vspace{-0.6cm}
			\includegraphics[width=5.2cm]{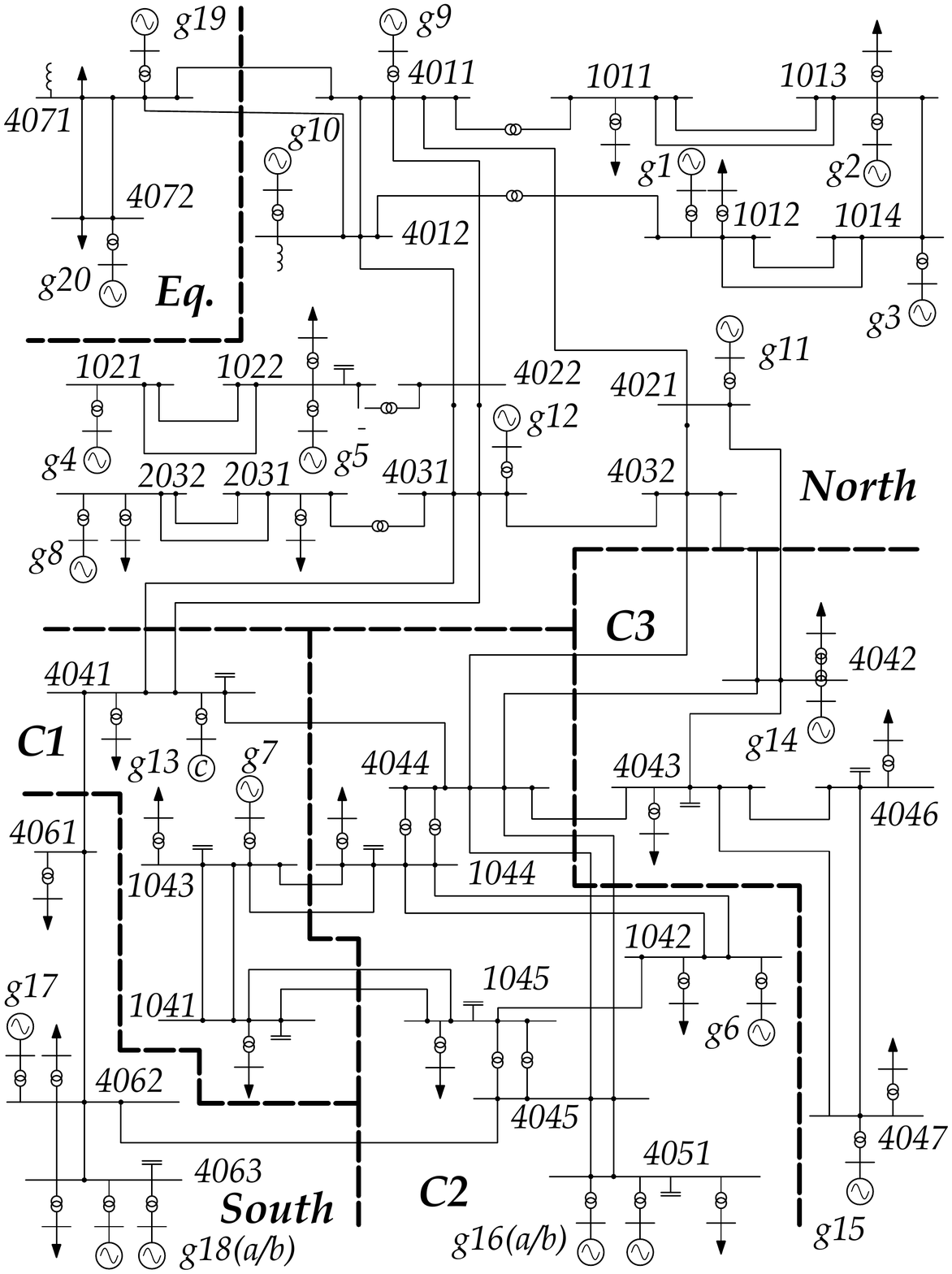}    
			\vspace{-0.6cm}
			\caption{One-line diagram of Nordic32 system with subareas} 
			\label{fig:Nordic32}
			\vspace{-0.7cm}
		\end{center}
	\end{figure}
	
	\subsection{Generation of training data}
	\label{sec:Gen}
\begin{figure}
		\begin{center}
			\vspace{-1.2cm}
			\includegraphics[width=6.2cm]{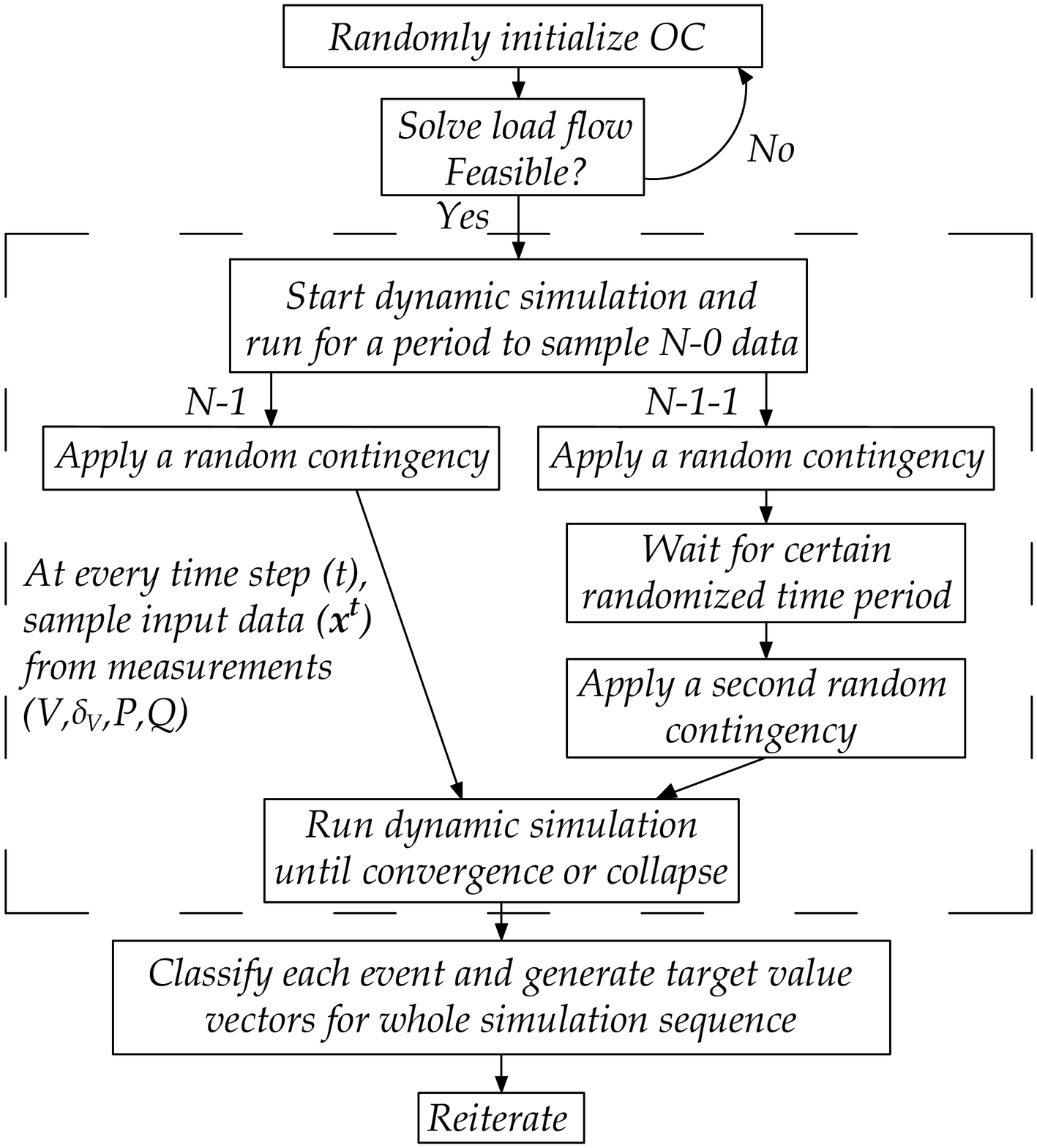}    
			\vspace{-1.3cm}
			\caption{Flowchart for generating input data and target values}
			
			\label{fig:LSTM_Flowchart}
			\vspace{-0.7cm}
		\end{center}
	\end{figure}

	The generation of a training set is a critical step, and a range of different initial OCs and contingencies were included to generate a representative training set. Dynamic simulations were performed using PSS®E 34.2.0 with its built-in models \cite{PSSEModel}. The steps of generating the training data are illustrated as a flowchart in Fig. \ref{fig:LSTM_Flowchart} and can be summarized as follows: 
	
	\subsubsection{Initial OCs}
	For the Nordic32 system, the initial OCs were randomly generated around the stable operation point denoted as "operating point B" in \cite{IEEETESTSYSTEM}. A large number of possible OCs were simulated by randomly initiating the loads from a uniform distribution around the base case load levels (80~\% of original load as lower limit and 120~\% as upper limit), while the power factor of the loads was kept constant. The total load change was distributed among the generators based on a weighted random distribution, where a higher rated capacity of a generator results in a higher probability to cover a larger share of the total load change. All generation that could not be supplied by the regular generators were distributed to the slack bus generator g20, see Fig. \ref{fig:Nordic32}. 
	
	In real applications, more delicate methods for efficient database generation and more careful generation of relevant OCs should be used \cite{Konstantelos2017,Thams2019}, where for instance the impact of unit commitment and topology changes are taken into account. 
	
	\subsubsection{Solve and check for feasibility}
	The generated OCs were solved with a power flow simulator, which served as a starting point for the dynamical simulation. If the system load flow did not converge, the initial OC was re-initialized. 
	
	\subsubsection{Start dynamic simulation and introduce contingencies}
	\label{sec:stabledata}
	
	Two separate dynamic simulations were then initiated for the $N$-$1$ and the $N$-$1$-$1$ cases. The process is illustrated in Fig.~\ref{fig:Classification_N-1-1}. For each of the two cases, the system runs without any contingencies for $65$ seconds to generate sufficient amount of $N$-$0$ data for the network to train on. At $t = 66$ seconds, the \textit{same} first contingency was applied to both of the cases. After an additional uniformly distributed random time periods in $[10 - 30]$ seconds after the first contingency, a secondary consecutive contingency was applied for the $N$-$1$-$1$ cases. Events resulting in several (near-)simultaneous contingencies were not taken into account ($N$-$k$ events). 
	
	The considered contingencies in the simulations were either (i) tripping of a generator, or (ii) a three-phased fault during $0.1$ seconds, followed by tripping the faulted line, which was then kept tripped during the remaining time of the simulation. The first contingency was chosen to be a major fault, meaning a fault on any transmission line connecting the different main areas in the system (excluding the "Eq." area, see Fig. \ref{fig:Nordic32}), \textit{or} any larger thermal generator in the "Central" area. The second contingency, for the $N$-$1$-$1$ cases, included tripping of \textit{any} transmission line in the whole system, excluding lines in the "Eq." area. No variations of load and generation were taken into account during the dynamic simulations as these, in the relatively short time period of the simulation, are presumed to have a small impact on the system stability. 
	
	\subsubsection{Sample inputs and run until stopping criteria}
	
	For each of the two cases, an input vector $\boldsymbol{x}^t$ consisting of measurements of all bus voltage magnitudes and angles, active and reactive power flows, were sampled every second and saved in a data file. No information regarding the type and location of applied the contingencies was sampled, as this information implicitly can be learned by the LSTM network. For instance, the LSTM network should be able to correlate a zero power flow in a transmission line with that line being out of service. 
	
	Each dynamic simulation ran for a total of $560$ seconds, but was, in the case of a major voltage collapse, stopped in advance. The simulation interval of $560$ seconds was chosen to allow time for \textit{all} dynamic events to occur and for the system to either stabilize or collapse. 
	
	\subsubsection{Classification}
	\label{sec:Classification}
	
	For each case, a sequence of true target value vectors $\boldsymbol{y}^1,...,\boldsymbol{y}^{560}$ was generated for every time step in the time-domain simulation. Each $\boldsymbol{y}^t$ in these sequences represents the classification of the system if the system is allowed to run from time $t$ up until $560$ seconds without any changes to the current system. As time progresses and new events occur, the class of $\boldsymbol{y}^t$ may change. The sequences consists of multidimensional vectors where the actual class is encoded using one-hot (binary) encoding. 
	
	The classification was performed according both to the severity and the location of the system degradation at the \textit{end} of the time-domain simulation. The system was defined as stable if \textit{all} transmission bus voltage magnitudes were above or equal to $1$ pu, in an alert state if \textit{any} transmission bus voltage magnitude ranged between $0.9<V<1.0 $  pu, and in an emergency state if \textit{any} transmission bus voltage magnitude was below $0.9$ pu. Overvoltages were not taken into account. 
	
	\begin{figure}
		\begin{center}\vspace{-2.3cm}
			\includegraphics[width=7.8cm]{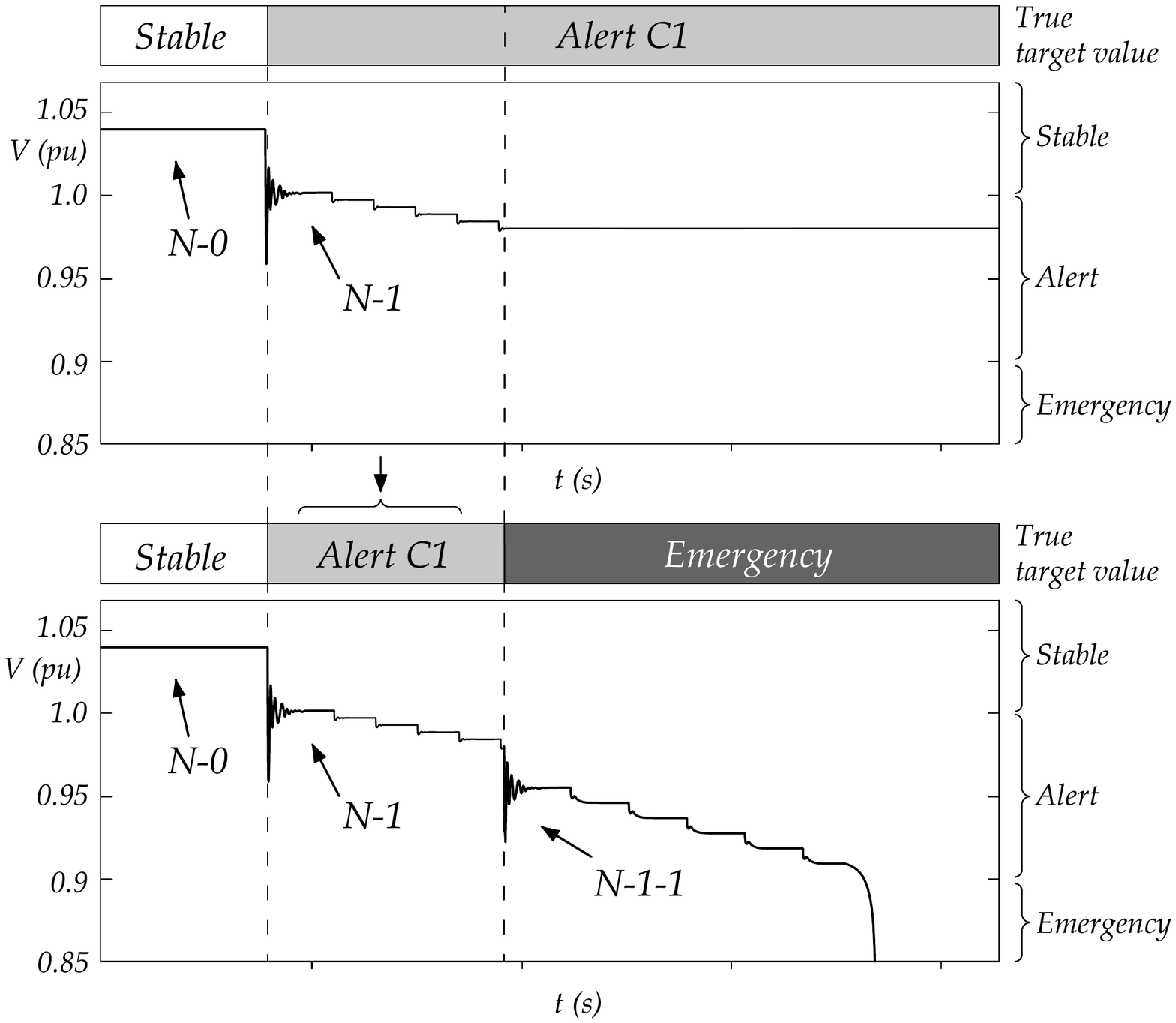}    
			\vspace{-2.5cm}
			\caption{Example of classification of an $N$-$1$ and an $N$-$1$-$1$ case} 
			\label{fig:Classification_N-1-1}
			\vspace{-0.8cm}
		\end{center}
	\end{figure}
	
	The target values for the \textit{alert} cases were also classified according to \textit{where} the lowest bus voltage magnitudes were found at the end of each dynamic simulation. The Nordic32 test system was therefore divided into different regions, as illustrated  in Fig. \ref{fig:Nordic32}. The regions "North", "South", and "Eq." were found to be stable regions, and no alert events were found in these regions for any of the simulated cases. Thus, for the classification of the alert cases, only the other three regions (indicated by \textbf{C1}, \textbf{C2}, \textbf{C3} in Fig. \ref{fig:Nordic32}) were used. The classification for each time step of each simulation belonged then to one of 5 different possibilities. Either, the whole system was predicted stable; it ended up in an emergency state; or an alert state was predicted in one of the three defined regions where the \textit{lowest} occurring transmission bus voltage was found.
	
	The classification process is illustrated in Fig. \ref{fig:Classification_N-1-1}. The target values are always classified as stable up until the first contingency. From different combinations of OCs and contingencies, the system may then end up being in a stable state, an alert state in area \textbf{C1}, \textbf{C2}, or \textbf{C3}, or in an emergency state. For the $N$-$1$ case, the sequence of true target value vectors from the time of the contingency to the end of the simulation are classified depending on which of these five states the systems ends up in. For the example of the $N$-$1$ case in Fig. \ref{fig:Classification_N-1-1}, the system ends up in an alert state in the \textbf{C1} area. For the $N$-$1$-$1$ case, the target values are classified as stable up until the first contingency. The target values are then gathered from the $N$-$1$ case, using the end state of that simulation for classifying the state \textit{between} the first and the consecutive contingency. After the second consecutive contingency, the system runs until it either collapses or until $560$ seconds. Depending on this final state, the sequence of true target value vectors from the second contingency until the end of the simulation are classified. In the example in Fig. \ref{fig:Classification_N-1-1}, an emergency state is reached. Note that the scales in the Fig. \ref{fig:Classification_N-1-1} are different from those in the simulations for easier interpretation. In real-life applications, more intricate stability limits could be used to allow a more detailed classification. 
	
	\subsubsection{Reiteration}
	The described steps are reiterated until a sufficiently large training set is generated. 
	
	\subsection{Architecture of the LSTM network}
\begin{figure}
		\begin{center}
			\vspace{-0.6cm}
			\includegraphics[width=7.0cm]{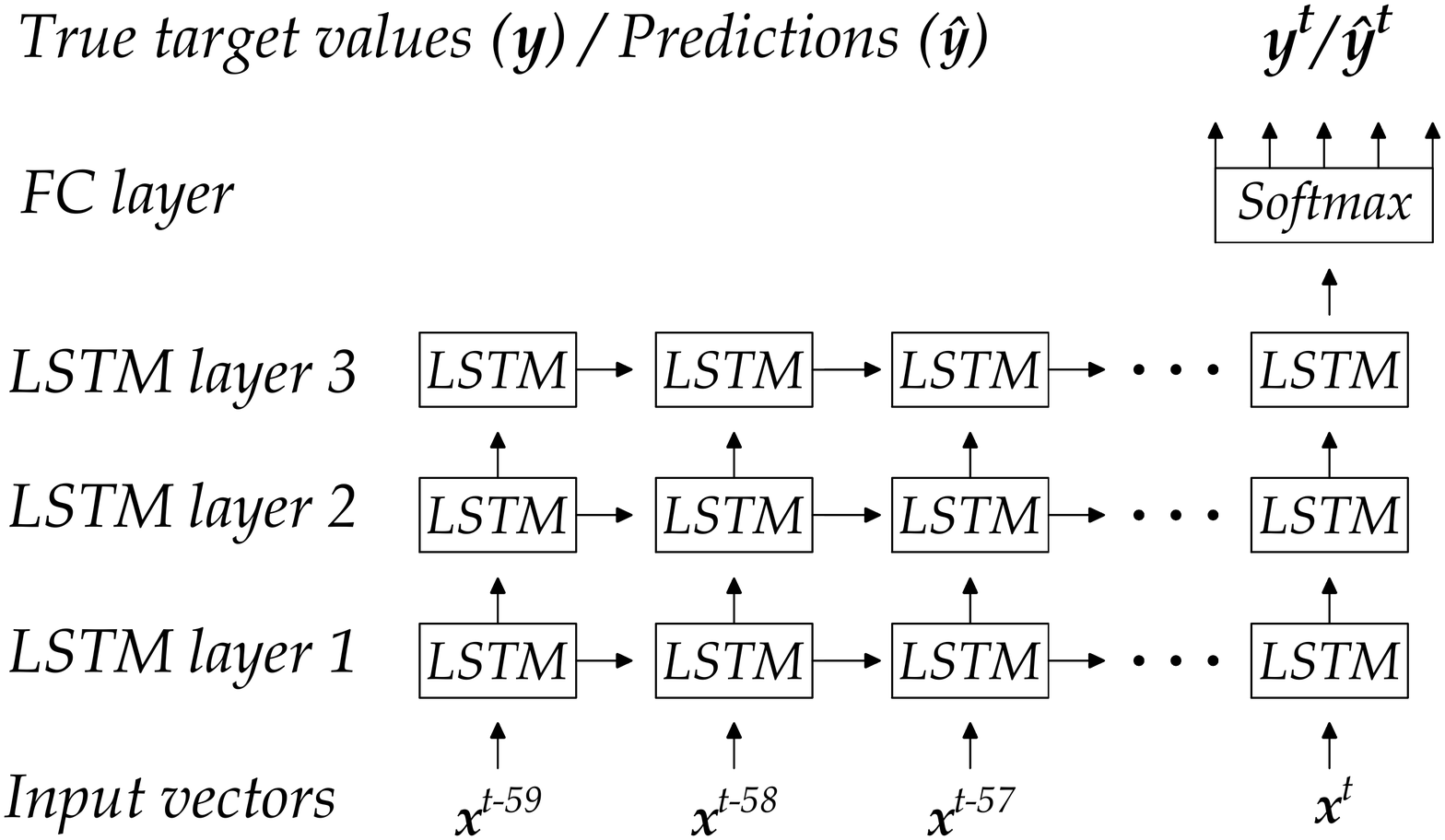}
			\vspace{-0.6cm}
			\caption{The proposed LSTM network architecture} 
			\label{fig:LSTM_overview}
			\vspace{-0.7cm}
		\end{center}
	\end{figure}

	The proposed LSTM network architecture, shown in Fig.~\ref{fig:LSTM_overview}, is generally referred to as a "many-to-one" architecture, where previous measurements in the time sequence are used for the classification in the final block. The network consists of three stacked LSTM layers which are used to capture different levels of features from the inputs. Each LSTM block consists of 32 individual LSTM cells. The first layer of LSTM-blocks takes a generated sequence of input vectors as inputs; then by mathematical operation as presented in Section \ref{sec:RNN}, the output of each block is forwarded both to the following block in the sequence, as well as to the upper layer of LSTM-blocks. The third layer of LSTM-blocks only passes the output forward along the time sequence. 
	The output layer at time $t$ is a fully connected network with softmax activation for classification. In training, the network use the true target vector $\boldsymbol{y}^t$ at time $t$, while during the test or prediction phase, the network estimates a prediction vector~$\hat{\boldsymbol{y}}^t$ at time $t$. The interpretation of the prediction problem is further explained in section \ref{sec:Interpretation}. 
	
	\subsection{Training the LSTM network}
	\label{sec:TrainingLSTM}
	
	Different data sets were used for training, validation, and testing of the method on a mix of $N$-$1$ and $N$-$1$-$1$ cases. The training data set has the dimension  $\left(135,000 \times 364 \times 560\right)$, where the dimension represents the number of training cases, the number of inputs, and the total interval in seconds for each simulation, respectively. 
	
	\begin{table}
		\caption{Design and hyperparameters used in training}
		\label{table:TrainingParameters}
		\vspace{-0.3cm}
		\centering
		\begin{tabular}{l l l}
			\toprule
			& \textbf{Parameter} & \textbf{Values and size}\\ 
			\midrule
			
			\multirow{5}{*}{{\rotatebox[origin=c]{90}{\textit{Data}}}} &
			Simulation interval &  560  \\
			& Feature dimension &  364  \\
			& Target classes &  5  \\
			& Training cases ($N$-$1$+$N$-$1$-$1$) & 45,000 + 90,000   \\
			& Validation cases ($N$-$1$/$N$-$1$-$1$) & 5,000 / 10,000 \\
			& Test cases ($N$-$1$/$N$-$1$-$1$) & 10,000 / 10,000 \\
			\addlinespace
			\multirow{5}{*}{{\rotatebox[origin=c]{90}{\textit{Architecture}}}} &
			
			LSTM layers &  3  \\
			& LSTM sequence length &  60  \\
			& FC activation function &  Softmax  \\
			& LSTM hidden cells & 32 \\
			& LSTM Activation function & Tanh\\
			
			\addlinespace
			\multirow{5}{*}{{\rotatebox[origin=c]{90}{\textit{Training}}}} &
			Max Epochs & 400  \\
			&Learning rate ($\alpha$) &  0.0001  \\
			& Dropout \& recurrent dropout &  50 \% / 50 \% \\
			&Optimizer &  Adam \cite{Kingma2014} \\
			&Loss metric &  Categorical cross-entropy\\

			\bottomrule
			\vspace{-0.8cm}
		\end{tabular}		
	\end{table}
	
	Before training, a process generally referred to as sequence preprocessing was performed to prepare batches of sequences with suitable length. The network is designed to take a sequence of $60$ time steps of measurements as inputs and subsequences with a length of $60$ time steps $\left(\boldsymbol{x}^{t-59},...,\boldsymbol{x}^{t}\right)$ were thus extracted from the $560$ seconds long simulation intervals, for different values of $t$. For each subsequence of input vectors, a corresponding target value $\left(\boldsymbol{y}^{t}\right)$ at time $t$ was gathered. The sequence preprocessing was performed $120$ times for \textit{each} training and validation case by varying $t$ between values of $t=[60,180]$. The lower bound of $t$ is required to always allow historic data to be included into the sequence. The LSTM network could have been trained on the whole simulation interval by increasing upper bound of $t$ from $180$ to $560$. However, since the method is proposed to be used in fast VIP applications, there is less usefulness of predicting instability long after the contingencies have occurred. 
	
	The generated subsequences were then used to train the LSTM network. Due to memory limitations, a method called mini-batch gradient descent was utilized where mini-batches of $1000$ subsequences were used separately to train the network. The training was performed for a maximum of 400 epochs. An epoch is finished when all generated batches have been used to update the network parameters. Adam \cite{Kingma2014}, an adaptable algorithm suitable for gradient-based optimization of stochastic objective functions was used in training the network. The algorithm used default parameters according to \cite{Kingma2014}, except for the learning rate which was tuned. The loss function which the optimizer is applied on is the categorical cross-entropy function, which is suitable for multi-classification problems. To avoid overfitting the data, two regularization techniques were used during the training. First, early stopping was implemented, and the training of the network was stopped in case the performance on the validation set did not improve after six epochs. Second, a technique called dropout was applied, where a certain percentage of the connections between inputs and the LSTM cells were randomly masked (or "dropped") with the aim of reducing overfitting on the data. Both conventional dropout and recurrent dropout between consecutive blocks were applied during the training phase. 
	
	All other parameters related to the training of the network are presented in Table \ref{table:TrainingParameters}. The architecture and parameters used to train the network have been iteratively tuned to increase the classification accuracy. However, the tuning could be extended even further to allow an even better classification accuracy.
	
	\subsection{Interpretation and intuition of the VIP problem}
	\label{sec:Interpretation}
	By the proposed training and architecture of the LSTM network, a classification problem is solved where the \textit{current} system state space is separated into different regions. Every state on a trajectory to a stable, alert (in \textbf{C1}, \textbf{C2}, or \textbf{C3}), or emergency state is labelled accordingly. The LSTM network is then trained on this data to implicitly learn these asymptotic properties of solutions and the trajectories of the system state. Once trained, the network can correlate the inputs, current and historic measurements, with a certain state space region and trajectory, allowing instant warnings of voltage instability only moments after a contingency have occurred in a system. The classification is performed under the assumption that the current system is unchanged, meaning that no additional contingencies or changes in generation and load configuration will occur. However, as time progresses, new observations are used as inputs to the LSTM network to continuously update and incorporate such changes in the system. 
	
	This VIP problem should be interpreted as a fixed horizon prediction problem, where the prediction horizon always is the final state given by the trajectories of the (dynamical) system. This interpretation assumes that the simulation horizon of the generated time-domain simulations are sufficiently long so that extending the simulation horizon even further, for this particular system beyond 560 seconds, would not change the partitioning of the state space. 
	
	\section{Results and discussion}
	\label{sec:Simulations}
	
	\subsection{Test results}
	
	\begin{table*}
		\vspace{-0.2cm}
		\caption{Confusion table showing prediction results and accuracy of the LSTM network evaluated at $T = 50$ seconds }
		\label{table:TrainingErrorsData}
		\vspace{-0.2cm}
		\centering
		\begin{tabular}[H]{l l l l l l l l l l l l l}
			\toprule
			& & & &  \multicolumn{7}{c}{\textit{Predicted states ($N$-$1$ / $N$-$1$-$1$)	}} & & \vspace{0.1cm}	\\ 
			&  & & & \textbf{Stable state}  & & \multicolumn{3}{c}{\textbf{Alert state}} & & \multicolumn{1}{c}{\textbf{Emergency state}} & & \textbf{Accuracy}	\vspace{0.1cm}	\\ 
			&  \textbf{Classification}   & & & All areas  & & C1 & C2 & C3 & & All areas  &  & \\ 
			
			\midrule
			\addlinespace
			\multirow{6}{*}{{\rotatebox[origin=c]{90}{\textit{Actual states}}}}  & \textbf{Stable state}   &  All areas    & &  2766 / 1171 &  & 0 / 20 & 0 / 19  & 0 / 0  &  & 0 / 0 & & 100 / 96.8~\%  \\
			\addlinespace
			&  \multirow{3}{*}{\textbf{Alert state}}  & C1  & & 0  /  0 &  & 856  /  565 & 0 / 5 & 0 / 0 &  & 0 / 0 & & 100 / 99.1~\%  \\ 
			&  & C2 && 0 / 5 && 0 / 8 & 1874 / 1237  & 0 / 1 &&  0 / 90  & & 100 / 92.2~\%  \\ 
			&  & C3 && 0 / 0 && 0 / 0 & 0 / 42 & 0 / 178 && 0 / 0 & & -~~~ / 89.9~\%  \\ 
			\addlinespace
			
			& 		\textbf{Emergency state} & All areas && 0 / 0 && 0 / 0 & 0 / 34 & 0 / 0  && 4504 / 6625 & & 100 / 99.5~\% \\
			\addlinespace \hline
			\addlinespace
			&  \textbf{Accuracy} & & & 100 / 99.6~\% && 100 / 95.3~\% & 100 / 92.5~\% & - / 99.4~\%  && 100 / 98.7~\% && \textbf{100 / 97.7~\%} \\
			\bottomrule
			\vspace{-0.6cm}
		\end{tabular}		
	\end{table*}
	
	The developed VIP methodology was tested on two separate test sets, one containing only $N$-$1$ cases, the other containing $N$-$1$-$1$ cases. Each test set was composed of $10,000$ cases of dynamic simulations. The test results of the predictions are presented using categorical accuracy, where the \textit{indices} of the true target values are compared to the argument maxima of the predictions. The accuracy at \textit{each} time step is then calculated over time for each of the two test sets. 
	
	The data were fed into the network in the form of a rolling window, with subsequences generated in the exact same manner as described in Section \ref{sec:TrainingLSTM}. As time $t$ progresses, new measurements entered the network from the rightmost block in the input layer and were shifted to the left in each time increment. Since the LSTM network require a sequence of 60 time steps of data, no predictions were made before $t=60$. To facilitate the presentation in the following figures, a new time index $T$ is introduced here. The relationship between the two time indices is $T=t-60$. The classification accuracy is only plotted for 120 seconds to better visualize the changes in accuracy after the contingencies.
	
\begin{figure}
		\begin{center}
			\vspace{-3.3cm}
			\includegraphics[width=7.8cm]{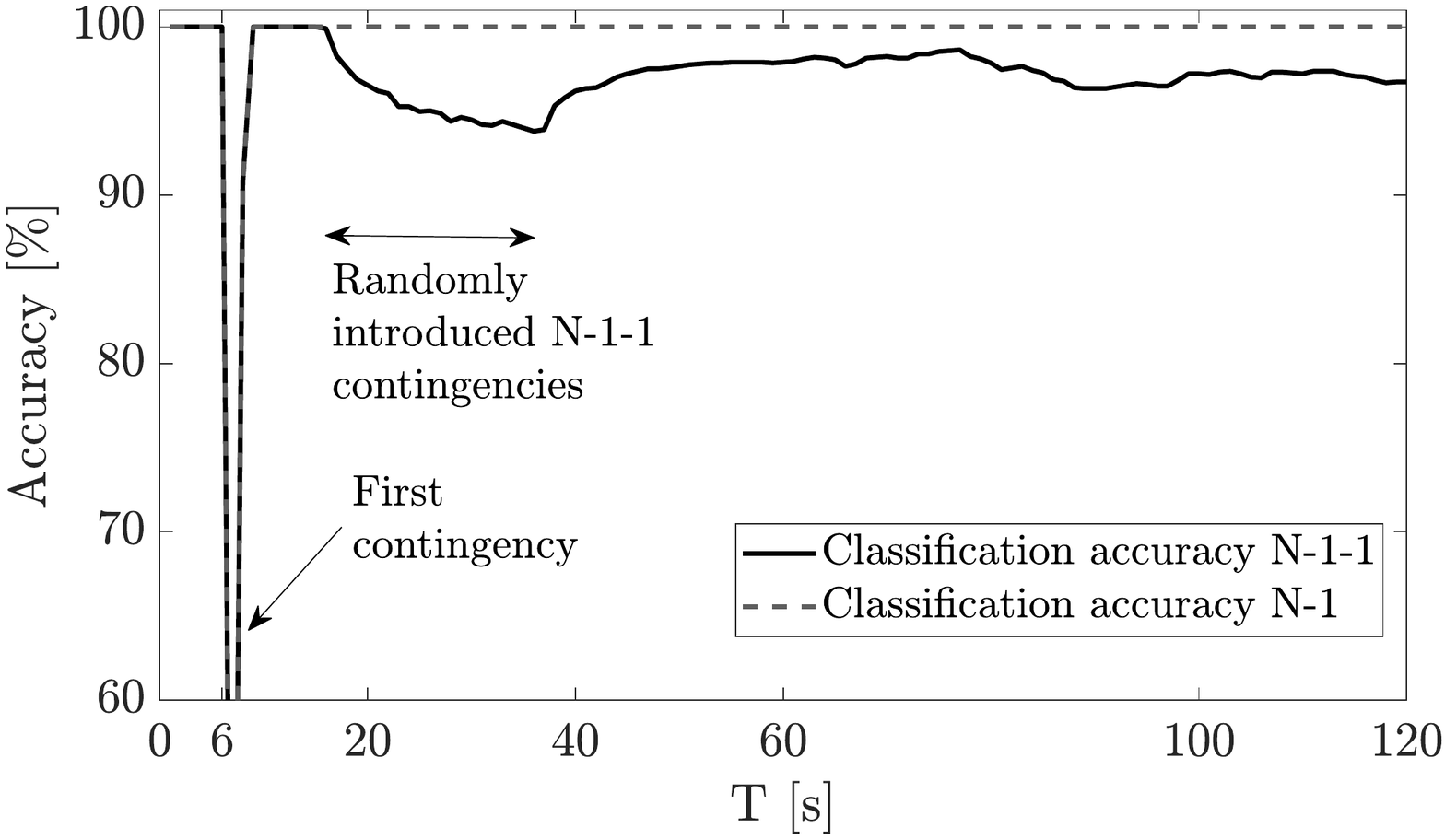}
			\vspace{-3.5cm}
			\caption{Classification accuracy over time for the proposed LSTM network} 
			\label{fig:Scores}
			\vspace{-0.8cm}
		\end{center}
	\end{figure}

	The classification accuracy \textit{over time} is presented in Fig.~\ref{fig:Scores}. The classification accuracy for the $N$-$1$ test set dropped significantly at $T = 6$ seconds, which is the same instant that the first contingency is applied. The large drop in classification accuracy can be attributed to low bus voltages instantaneously following the first contingency, which the LSTM network has learned to correlate to a voltage instability event. After the first contingency, the classification accuracy increased and remained constant at 100~\% for the rest of the simulations. 
	
	The classification accuracy for the $N$-$1$-$1$ test set was identical up until the consecutive contingencies were randomly applied. During this time span, illustrated by the arrows in Fig.~ \ref{fig:Scores}, the classification accuracy decreased slightly. Since these contingencies do not occur at the same time instant in each test case, the same large drop in accuracy as for the $N$-$1$ cases was not seen. The accuracy then gradually increased and stabilized at around $96$-$98~\%$. 
	
	The results show that the LSTM network can classify and predict future stability almost perfectly for the N-1 contingency cases and with good accuracy for the N-1-1 cases. To examine which cases were misclassified, the prediction accuracy for the two test sets, evaluated at $T=50$ seconds, are presented in Table \ref{table:TrainingErrorsData} in the form of a confusion table. Each number in the column in the table represents instances of the predicted classes and each number in the row represents the instances of the actual classes. The (empirical) conditional probabilities of correctly classifying a certain state is presented in the column furthest to the right. Similarly, the conditional probability of a state \textit{actually} belonging to the predicted state is presented in the bottom row of the table. The total accuracy is presented in the lower right corner of the table. The accuracy for all $N$-$1$ cases is 100 \% and no cases are falsely classified. For the $N$-$1$-$1$ test set, the lowest classification accuracy occurred for the alert states. After inspection of the falsely classified cases, it was found that several of these were borderline cases where the transmission bus voltage magnitude used in the classification were very close to what was used in the other classes. The highest classification accuracy occurred for the emergency cases with $99.5~\%$. 
	
	It should be noted that the test and training sets were weighted with more cases ending up in certain classes than others. It is thus probable that the results are slightly biased with higher accuracy for these classes, and that the classification accuracy of the other classes may be lower as an effect.

	\subsection{Impact of sequence length}
	\label{sec:sequencelength}
	
	In this section, the performance of the sequence-based approach is tested and compared against a conventional feedforward NN, which only uses a single snapshot of measurements as inputs. Further, to test the impact of a \textit{shorter} time sequence, the results of an LSTM network using a time sequence of $30$ time steps, instead of $60$, are presented. 
	
	To allow a fair comparison between the two approaches, the feedforward NN used in this comparison was designed to be as similar as possible to the LSTM network. Essentially, the design of the NN in the comparison is identical to the \textit{final} time step in the LSTM network presented in Fig. \ref{fig:LSTM_overview}, with the difference that each layer consists of a hidden layer of neurons. The designed NN thus has three hidden unit layers, each layer with 32 hidden nodes. The same FC layer with a softmax activation function was used. The training for the NN was performed identically as for the LSTM network, with the exception that instead of a sequence of input values, a single snapshot was used. The LSTM network using a shorter time sequence was trained identically to that of the longer LSTM network with the exception that a shorter sequence of $30$ instead of $60$ time steps was used. 
	
\begin{figure}
		\begin{center}
			\vspace{-2.8cm}
			\includegraphics[width=6.6cm]{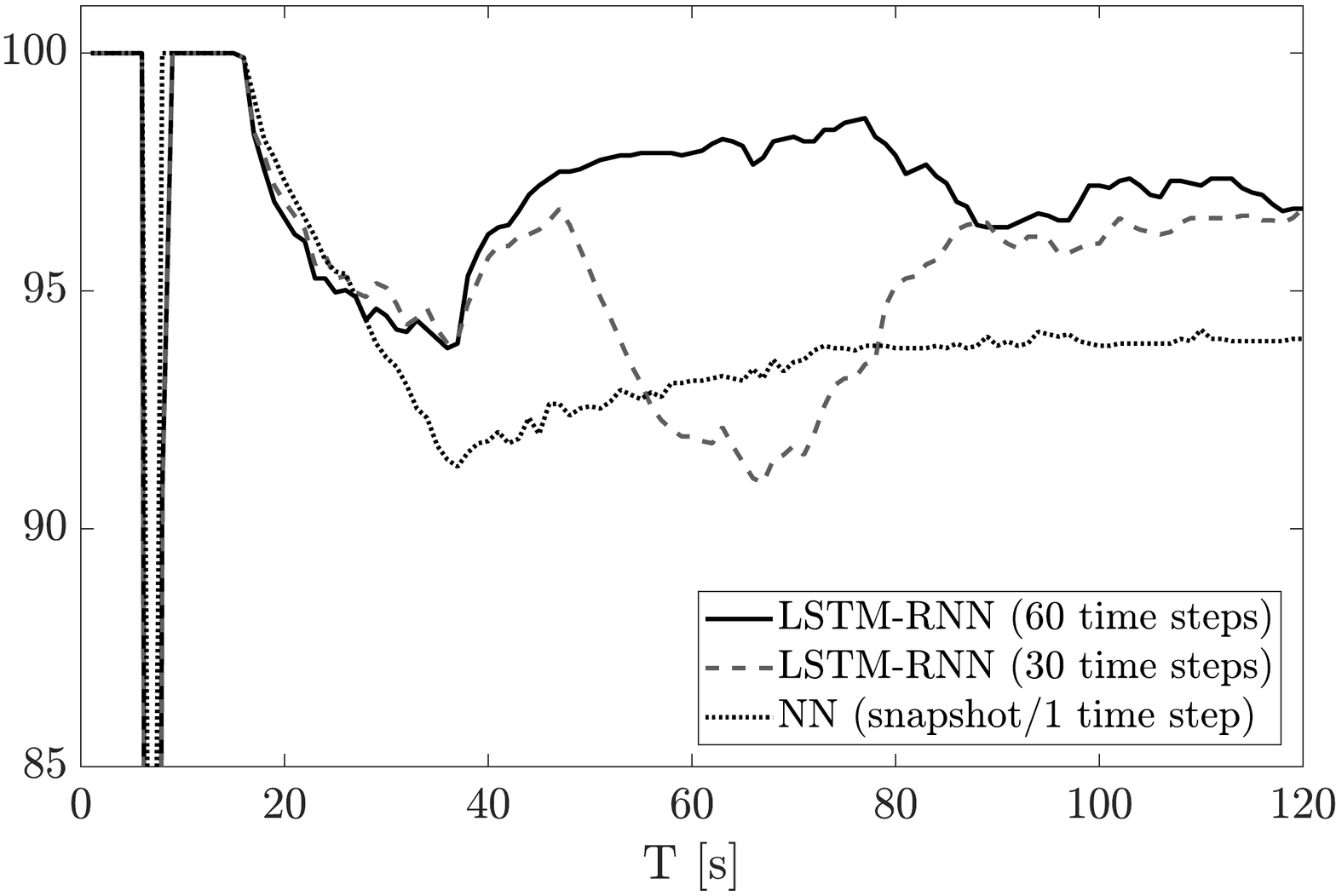}
			\vspace{-2.8cm}
			\caption{Impact of sequence length on classification accuracy} 
			\label{fig:Time_seq_comp}
			\vspace{-0.7cm}
		\end{center}
	\end{figure}

	In Fig. \ref{fig:Time_seq_comp}, the classification accuracy on the $N$-$1$-$1$ test set is presented for the two LSTM networks with the different time sequence length and for the conventional NN. The classification accuracy for the conventional NN was around $93~\%$ after all the consecutive contingencies were been applied, while that of the proposed LSTM network is around $96$-$98~\%$. The results clearly indicate that the performance of the LSTM network using $60$ time steps in the sequence significantly exceeded that of the conventional NN, generally providing better classification accuracy over the whole time frame of the simulation cases. 
	
	The classification accuracy of the LSTM network using a shorter sequence was similar to the one using a longer sequence, with the difference of a large drop in classification accuracy at around $T$ = $46$ seconds, see Fig. \ref{fig:Time_seq_comp}. The same decline in classification accuracy, though less significant, can be noted for the LSTM network using the longer time sequence. For the LSTM network with the $60$ time steps long sequence, the accuracy drop started at $T = 76$ seconds, declined for $20$ seconds, and was then restored to around $97~\%$ accuracy. For the LSTM network using the $30$ time steps long sequence, the decline started $30$ time steps earlier, at $T$ = $46$ seconds. Once again, the classification accuracy decreased for $20$ seconds, and was then restored.

	One explanation of these results is that the LSTM networks utilize information concerning the contingency and \textit{pre-contingency} state to enhance the classification accuracy. It should be noted that the decline in classification accuracy started exactly $60$ respectively $30$ seconds after the consecutive contingencies are introduced (at $T=16$) and that the duration for the decline in classification accuracy corresponded to the exact time frame that the consecutive contingencies were introduced. Thus, when the networks lose the information about the pre-contingency state, the chance of a misclassification increases. These results strengthen the hypothesis a long sequence LSTM network could be used to enhance the state signal to provide better classification accuracy. Theoretically, an even longer sequence could be used to increase the accuracy even further. However, this would increase the computational cost of training, and a balance between classification accuracy and computational cost should be sought. 
	
	\subsection{Generalization capability and training set requirement}
	
	The generalization capability of a ML method refers to the capability to generalize the learning from the actual training set to other, yet unseen cases.
	Such capability is especially valuable in overcoming the combinatorial increase of complexity in the training when $N$-$1$-$1$ cases are also considered~\cite{Mitra2016}.
	
\begin{figure}
		\begin{center}
			\vspace{-2.8cm}
			\includegraphics[width=6.6cm]{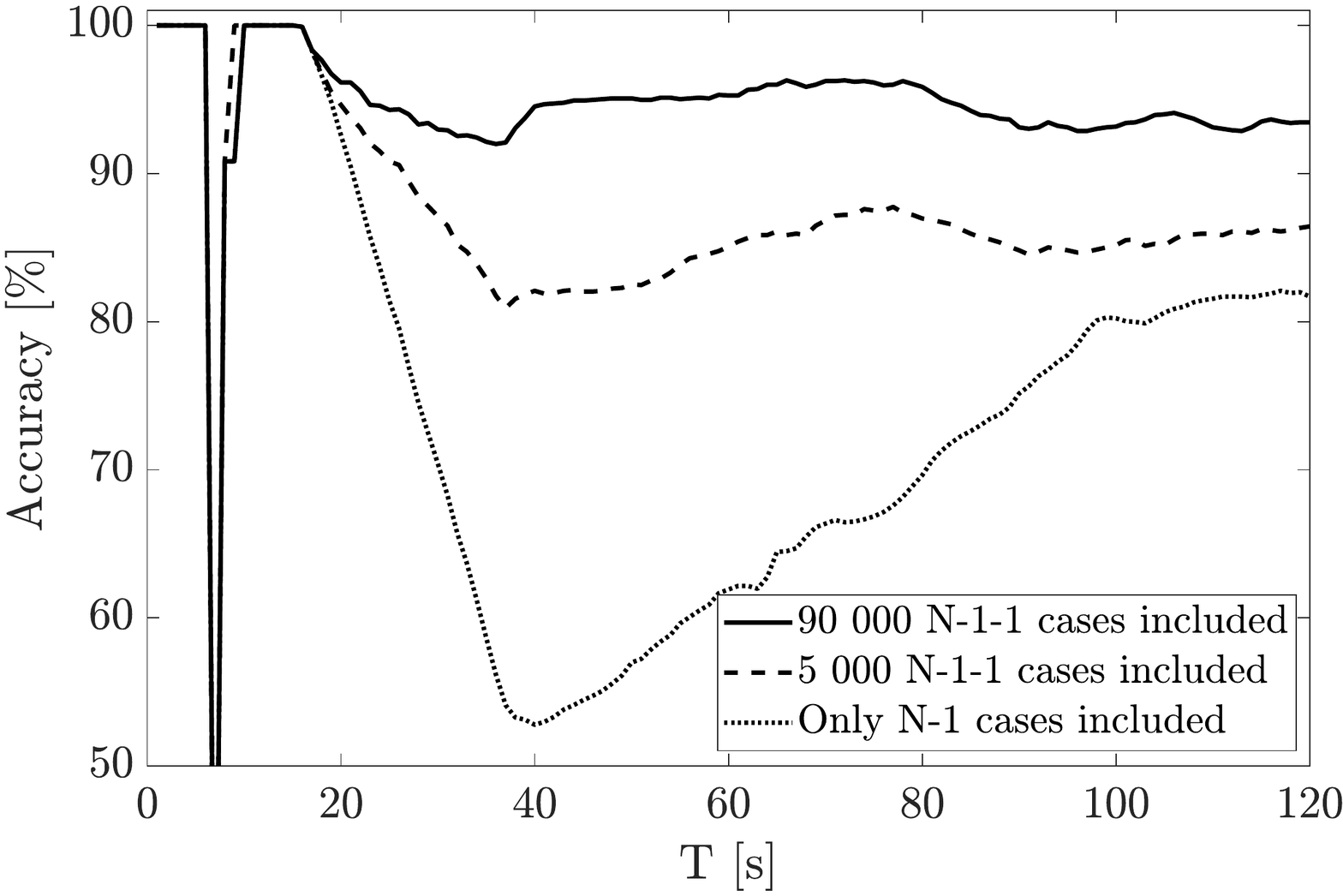}
			\vspace{-2.8cm}
			\caption{Classification accuracy over time when varying the number of $N$-$1$-$1$ cases included in the training data}
			\label{fig:Scores_TestN-1}
			\vspace{-0.7cm}
		\end{center}
	\end{figure}

	In Fig. \ref{fig:Scores_TestN-1}, the classification accuracy is presented on the $N$-$1$-$1$ test set when the LSTM network have been trained on three different training sets. The results are presented when the network were trained on i) the full training set with all $N$-$1$ and $N$-$1$-$1$ cases included, ii) a smaller training set with all $N$-$1$ cases but where only a small batch ($5,000$) of $N$-$1$-$1$ cases have been included, and iii) a training set where the network is \textit{only} trained on $N$-$1$. The same training approach as previously described were used. According to Fig. \ref{fig:Scores_TestN-1}, the classification accuracy was significantly reduced when no $N$-$1$-$1$ cases are included in the test set. When including the small batch ($5,000$) of $N$-$1$-$1$ cases, the classification accuracy increased significantly. However, the accuracy is still lower than when the full training set is used. Thus, the importance of obtaining a representative training set is still imperative if a high classification accuracy is to be achieved. 
	
	\subsection{Practical applications and requirements}
	\label{sec:Applications}
	The method is proposed to be used as an online tool for system operators to monitor the current state of a power system. It should be stressed that the method is not proposed to replace conventional voltage instability detection methods, but rather function as a supplementary tool to provide early warnings. The instantaneous prediction capability of the proposed method has to be weighed against the possibility of misclassification of the system's future stability. When comparing the proposed method to other conventional indicators for voltage instability detection (see \cite{Glavic2011}), it is important to remember that these might be more accurate once instability detected, but generally take \textit{significantly} longer time to indicate instability, thus reducing the time frame that system operators have to steer the system back into stable operation. 
	
	For the proposed method to be effective, measurement updates should be available within a few seconds. In this paper, a measurement update rate each second is assumed to be available. To assure that errors and missing values are filtered out, measurements should always be preceded by a state estimator. However, state estimates from a non-linear state estimator based on remote terminal units may be too slow to be effective. Thus, time-synchronized measurements from wide-area phasor measurements filtered through a linear state estimator would be preferred. 
	
	The softmax classifier of the LSTM network outputs a probability vector, where each class is given a certain probability. It should be noted that this probability vector does not provide a \textit{true} representation of the model confidence. However, it can still be useful as a proxy by system operators to track the network's confidence in each prediction. Thus, the operator can use the probability vector directly in an online interface to track the network's belief in each prediction. Alternatively, argument maxima or other functions could be used to present the most probable prediction of the network, or, for instance, to avoid predictions of falsely labelled stable states. 
	
	The \textit{practical} classification accuracy of the proposed method will be affected by many aspects and will generally be lower than on a simulated test set. One of the more important aspects are modeling errors, including erroneous system parameters or inaccurate modeling of parameter values for dynamic models. Such aspects will introduce a difference between the simulated and the actual dynamic response after a contingency. However, it should be noted that such limitations are not limited only to ML based approaches for VIP. All methods for DSA require that the dynamic models used in assessing the system response are accurately modeled. 
	
	\section{Conclusions}
	\label{sec:Conclusions}
	
	This paper presents a new approach for online voltage instability prediction using an LSTM network capable of utilizing a sequence of measurements to improve classification accuracy. Once trained, the LSTM network can allow system operators to continuously assess and predict whether the present system state is stable, or will evolve into an alert or an emergency state in the near future. The network is also adapted to be able to indicate \textit{where} instability emerges, allowing system operators to perform more cost-effective control measures. 
	
	The LSTM network was proposed with the aim of improving the available state signal by implicitly learning the long-term dependencies of voltage instability events. The results presented in the paper are highly encouraging and the proposed method is shown to have high accuracy in predicting voltage instability. The impact of sequence length of the LSTM network was tested and showed that a longer sequence provided a significantly better classification capability than both a feedforward NN and a network using a shorter sequence. The paper also examined the generalization capability of the proposed LSTM network, where the classification accuracy on $N$-$1$-$1$ cases was assessed when the system was only trained on $N$-$1$ cases. It was found that this reduced the classification accuracy significantly, whereas including a smaller subset of $N$-$1$-$1$ cases into the training set resulted in significantly better performance. 
	

	\ifCLASSOPTIONcaptionsoff
	\newpage
	\fi

	
	
	%
	\bibliography{Referenser}

\begin{thebibliography}{10}
\providecommand{\url}[1]{#1}
\csname url@samestyle\endcsname
\providecommand{\newblock}{\relax}
\providecommand{\bibinfo}[2]{#2}
\providecommand{\BIBentrySTDinterwordspacing}{\spaceskip=0pt\relax}
\providecommand{\BIBentryALTinterwordstretchfactor}{4}
\providecommand{\BIBentryALTinterwordspacing}{\spaceskip=\fontdimen2\font plus
\BIBentryALTinterwordstretchfactor\fontdimen3\font minus
  \fontdimen4\font\relax}
\providecommand{\BIBforeignlanguage}[2]{{%
\expandafter\ifx\csname l@#1\endcsname\relax
\typeout{** WARNING: IEEEtran.bst: No hyphenation pattern has been}%
\typeout{** loaded for the language `#1'. Using the pattern for}%
\typeout{** the default language instead.}%
\else
\language=\csname l@#1\endcsname
\fi
#2}}
\providecommand{\BIBdecl}{\relax}
\BIBdecl

\bibitem{IEEE2004}
P.~Kundur \emph{et~al.}, ``{Definition and classification of power system
  stability IEEE/CIGRE joint task force on stability terms and definitions},''
  \emph{IEEE Trans. Power Syst.}, vol.~19, no.~3, pp. 1387--1401, Aug 2004.

\bibitem{Glavic2011}
M.~Glavic and T.~Van~Cutsem, ``A short survey of methods for voltage
  instability detection,'' in \emph{Proc. (IEEE) PES General Meeting}, Detroit,
  MI, Jul 2011, pp. 1--8.

\bibitem{Konstantelos2017}
I.~Konstantelos \emph{et~al.}, ``Implementation of a massively parallel dynamic
  security assessment platform for large-scale grids,'' \emph{IEEE Trans. Smart
  Grid}, vol.~8, no.~3, pp. 1417--1426, May 2017.

\bibitem{Cutsem1993}
T.~Van~Cutsem \emph{et~al.}, ``Decision tree approaches to voltage security
  assessment,'' \emph{IEE Proceedings C - Generation, Transmission and
  Distribution}, vol. 140, no.~3, pp. 189--198, May 1993.

\bibitem{Mansour1997}
Y.~Mansour \emph{et~al.}, ``Large scale dynamic security screening and ranking
  using neural networks,'' \emph{IEEE Trans. Power Syst.}, vol.~12, no.~2, pp.
  954--960, May 1997.

\bibitem{Sun2007}
K.~Sun \emph{et~al.}, ``An online dynamic security assessment scheme using
  phasor measurements and decision trees,'' \emph{IEEE Trans. Power Syst.},
  vol.~22, no.~4, pp. 1935--1943, Nov 2007.

\bibitem{Khoshkhoo2014}
H.~Khoshkhoo and S.~M. Shahrtash, ``Fast online dynamic voltage instability
  prediction and voltage stability classification,'' \emph{IET Generation,
  Transmission \& Distribution}, vol.~8, no.~5, pp. 957--965, May 2014.

\bibitem{Liu2018}
C.~Liu, F.~Tang, and C.~L. Bak, ``An accurate online dynamic security
  assessment scheme based on random forest,'' \emph{Energies}, vol.~11, no.~7,
  2018.

\bibitem{Diao2009}
R.~Diao \emph{et~al.}, ``{Decision tree-based online voltage security
  assessment using PMU measurements},'' \emph{IEEE Trans. on Power Syst.},
  vol.~24, no.~2, pp. 832--839, May 2009.

\bibitem{Khoshkhoo2011}
H.~Khoshkhoo and S.~M. Shahrtash, ``On-line dynamic voltage instability
  prediction based on decision tree supported by a wide-area measurement
  system,'' \emph{IET Generation, Transmission \& Distribution}, vol.~6,
  no.~11, pp. 1143--1152, November 2012.

\bibitem{Hagmar2019}
H.~Hagmar \emph{et~al.}, ``On-line voltage instability prediction using an
  artificial neural network,'' in \emph{2019 IEEE Milano PowerTech}, June 2019
  (Accepted), pp. 1--6.

\bibitem{Sutton2015}
R.~S. Sutton and A.~G. Barto, \emph{Reinforcement Learning: An
  Introduction}.\hskip 1em plus 0.5em minus 0.4em\relax Cambridge,
  Massachusetts London, England: The MIT Press, 2015.

\bibitem{Schmidhuber2017}
K.~Greff \emph{et~al.}, ``{LSTM: A Search Space Odyssey},'' \emph{IEEE Trans.
  Neural Netw. \& Learning Syst.}, vol.~28, no.~10, pp. 2222--2232, Oct 2017.

\bibitem{Rumelhart1986}
D.~E. Rumelhart, G.~E. Hinton, and R.~J. Williams, ``Learning representations
  by back-propagating errors,'' \emph{Nature}, vol. 323, no. 6088, p. 533,
  1986.

\bibitem{Hochreiter1997}
S.~Hochreiter and J.~Schmidhuber, ``Long short-term memory,'' \emph{Neural
  computation}, vol.~9, pp. 1735--80, 12 1997.

\bibitem{IEEETESTSYSTEM}
\BIBentryALTinterwordspacing
T.~Van~Cutsem \emph{et~al.}, ``Test systems for voltage stability analysis and
  security assessment,'' IEEE/PES Task Force, Tech. Rep. PES-TR19, Aug. 2015.
  [Online]. Available:
  \url{http://resourcecenter.ieee-pes.org/pes/product/technical-publications/PESTR19}
\BIBentrySTDinterwordspacing

\bibitem{PSSEModel}
\emph{PSS®E 34.2.0 Model Library}, Siemens Power Technologies International,
  Schenectady, NY, Apr. 2017.

\bibitem{Thams2019}
F.~Thams \emph{et~al.}, ``Efficient database generation for data-driven
  security assessment of power systems,'' \emph{IEEE Trans. Power Syst.}, pp.
  1--1, 2019.

\bibitem{Kingma2014}
D.~P. {Kingma} and J.~{Ba}, ``{Adam: A Method for Stochastic Optimization},''
  \emph{arXiv e-prints}, p. arXiv:1412.6980, Dec 2014.

\bibitem{Mitra2016}
P.~Mitra \emph{et~al.}, ``A systematic approach to ${n}$-$1$-$1$ analysis for
  power system security assessment,'' \emph{IEEE Power and Energy Technology
  Systems Journal}, vol.~3, no.~2, pp. 71--80, June 2016.

\end{thebibliography}

	%
	
	\begin{IEEEbiographynophoto}{Hannes Hagmar}
		(S'17) received the M.Sc. degree in electric power engineering from Chalmers University of Technology, Gothenburg, Sweden in 2016. Between 2016 to 2017, he worked at RISE Research Institutes of Sweden with research in electric transmission systems and measurement technology. He is currently pursuing the PhD degree at Chalmers University of Technology. His research interest includes power system dynamics and stability, integration of renewables, and machine learning. 
	\end{IEEEbiographynophoto}
	
	\begin{IEEEbiographynophoto}{Lang Tong}
		Lang Tong (S'87,M'91,SM'01,F'05) is the Irwin and Joan Jacobs Professor of Engineering at Cornell University and the Cornell site Director of Power Systems Engineering Research Center (PSERC). He received the B.E. degree from Tsinghua University and the Ph.D. degrees in electrical engineering from the University of Notre Dame. He was a Postdoctoral Research Affiliate at the Information Systems Laboratory, Stanford University held visiting positions at Stanford University, the University of California at Berkeley, the Delft University of Technology, and the Chalmers University of Technology in Sweden.   Lang Tong's current research focuses on optimization, machine learning, AI, and economic problems in energy and power systems. He received several IEEE society transaction prize papers and conference best paper awards. He was a Distinguished Lecturer of the IEEE Signal Processing Society and the 2018 Fulbright Distinguished Chair in Alternative Energy.
	\end{IEEEbiographynophoto}
	
	\begin{IEEEbiographynophoto}{Robert Eriksson}
		(SM’16) received the M.Sc. and Ph.D. degrees in electrical engineering from the KTH Royal Institute of Technology, Stockholm, Sweden, in 2005 and 2011, respectively. He held an associate professor position at the Center for Electric Power and Energy, DTU Technical University of Denmark, from 2013 to 2015. He is currently with the Swedish National Grid, Department of Markets and System Development. His current research interests include power system dynamics and stability, automatic control, HVDC systems, and DC grids.
	\end{IEEEbiographynophoto}

	\begin{IEEEbiographynophoto}{Le Anh Tuan}
		(S’01–M’09) received the M.Sc. degree in energy economics from the Asian Institute of Technology, Bangkok, Thailand, in 1997, and the Ph.D.  degree  in  power systems  from the Chalmers University of Technology, Gothenburg, Sweden, in 2004. He is currently a Senior Lecturer with the Division of Electric Power Engineering, Department of Energy and Environment, Chalmers University of Technology. His current research interests include power system operation and planning, power market and deregulation issues, grid integration of renewable energy, and plug-in electric vehicles.
	\end{IEEEbiographynophoto}

\end{document}